\documentclass[review,authoryear,3p,times,12pt]{elsarticle}

\usepackage{multirow,setspace,times,amssymb,amsmath,graphicx,color,rotating,subfigure,url}
\usepackage{lineno,color}
\usepackage{natbib}
\usepackage{booktabs}%
\usepackage{longtable}%
\usepackage{rotating}
\usepackage{lscape}
\usepackage{threeparttable}
\graphicspath{{Figures/}}
\usepackage[table]{xcolor}
\usepackage{tabularx}
\usepackage{graphicx,color} 
\usepackage{epstopdf}
\usepackage{mathrsfs}
\usepackage{makecell}
\usepackage[bookmarks=true,colorlinks,linkcolor=blue,anchorcolor=blue,citecolor=blue,unicode,urlcolor=blue]{hyperref}
\usepackage{bookmark}
\usepackage{todonotes}
\usepackage{ragged2e}
\usepackage[font=small]{caption}
\usepackage{multirow}
\usepackage{enumerate}
\usepackage{afterpage}
\usepackage{dcolumn}

\hypersetup{CJKbookmarks=true}%
\journal{}

\begin{document}
\captionsetup[figure]{labelfont={bf},name={Fig.},labelsep=period}
\captionsetup[table]{labelfont={bf},name={Table},labelsep=newline,singlelinecheck=false}

\begin{frontmatter}

\title{From centrality to productivity: How firms reconfigure technological search in innovation networks?}
\author[HNU]{Han-Yun Tu}
\author[HNU]{Xiang Yang\corref{cor1}}
\ead{youngxianglove@163.com} 
\cortext[cor1]{Corresponding author.}
\author[ECUST,RCE]{Si-Yao Wei}
\address[HNU]{School of International Business, Hainan University, Haikou 570228, China}
\address[ECUST]{School of Business, East China University of Science and Technology, Shanghai 200237, China}
\address[RCE]{Research Center for Econophysics, East China University of Science and Technology, Shanghai 200237, China}

\begin{abstract}
Firms' positions in innovation networks determine their access to external knowledge, yet how these positions shape technological search behavior and influence productivity remains underexplored. We propose that central network positions systematically reconfigure firms' innovation strategies by promoting exploratory search across emerging technological domains while sustaining broader technological portfolios. This behavioral reorientation allows central firms to diversify their innovation efforts and leverage knowledge spillovers more effectively, translating network advantages into higher productivity. Using panel data on Chinese listed firms and patent-based measures of innovation networks, we construct a dynamic patent citation network to track changes in firms' network centrality and technological search patterns over time. Our findings show that firms with greater centrality are more likely to enter novel technological fields and expand their technological scope, leading to measurable gains in total factor productivity. We further demonstrate that the impact of network centrality on exploratory search is amplified by scientific embeddedness, whereas the productivity returns from exploration depend on technological distance. By connecting structural network positions with behavioral adaptations in technological search, this study uncovers a direct micro-level mechanism through which innovation networks drive firm performance. These results highlight the strategic value of network centrality in shaping not just access to knowledge, but also the direction and efficiency of innovation activities.
\end{abstract}

\end{frontmatter}


\section{Introduction}
\label{Sec_Introduction}

Sustained productivity growth remains a central, yet increasingly complex, challenge for innovative firms. While the resource-based view emphasizes that competitive advantage arises from distinctive technological capabilities and heterogeneous knowledge combinations \citep{FJ-Arts-2023-MS,FJ-Brennecke-2017-RP}, firms' technological evolution is frequently constrained by path-dependent search processes that generate core rigidities \citep{FJ-Benner-2003-AMR,FJ-Quintana-2008-RP,FJ-Stuart-1996-SMJ}. Such rigidity can trap firms in a ``success trap'', where past successes reinforce continued investment in mature technological domains while discouraging exploration of emerging opportunities. However, firms' innovation activities are embedded in broader knowledge networks that shape the accessibility of technological opportunities and influence the direction of technological search \citep{FJ-Guan-2016-RP,FJ-Lyu-2019-TFSC}. Understanding how firms translate their positions in these networks into sustained productivity growth therefore represents an important challenge in innovation research.

Existing studies have made substantial progress in examining the role of network positions in shaping firm behavior and outcomes. One stream of research focuses on the structural properties of innovation or alliance networks and shows that central positions provide privileged access to diverse knowledge and resources, which can enhance firm performance \citep{FJ-Ahuja-2000-ASQ,FJ-Chuluun-2017-JCF,FJ-Lyu-2019-TFSC,FJ-Tsai-2001-AMJ,FJ-Van-2015-RP,FJ-Zaheer-2005-SMJ,FJ-Zhang-2020-TFSC}. A second stream emphasizes the behavioral foundations of technological change and investigates how firms allocate search efforts between exploration and exploitation across technological domains \citep{FJ-Fleming-2001-MS,FJ-Gilsing-2008-RP,FJ-Lyu-2019-TFSC,FJ-March-1991-OS,FJ-Wen-2021-T,FJ-Zhong-2021-AMJ}. However, these two streams have largely evolved in parallel. Studies of innovation networks often show that firms occupying central positions tend to achieve superior performance \citep{FJ-Kao-2019-IJPE}, yet they rarely examine the behavioral mechanisms through which these structural advantages influence firms' innovation activities. In contrast, research on technological search highlights how firms allocate innovation efforts across exploration and recombination, but often pays limited attention to how external network structures shape these search processes. For example, \cite{FJ-Gilsing-2008-RP} found that in alliance networks, the interaction between technological distance and betweenness centrality can reduce exploratory innovation due to cognitive overload, while \cite{FJ-Lyu-2019-TFSC} used patent citation networks and PageRank centrality to demonstrate links to innovation outcomes, they rarely trace the behavioral pathways. This indicates a critical gap: we still lack an integrated understanding of how network centrality shapes firm behavior and, through this behavior, influences productivity.

In this study we argue that central positions in innovation networks reshape firms' technological search patterns, thereby influencing productivity outcomes. Exposure to heterogeneous knowledge signals circulating within innovation networks encourages firms occupying central positions to reconfigure their technological search \citep{FJ-Bianchi-2011-T,FJ-Lyu-2019-TFSC}. Rather than concentrating innovation efforts within established knowledge trajectories, central firms are more likely to redirect search toward emerging technological domains while maintaining broader technological portfolios. This reconfiguration of technological search may reduce firms' overreliance on established technological trajectories and improve the efficiency with which innovation resources are allocated across evolving technological opportunities \citep{FJ-Benner-2003-AMR}. At the same time, the productivity implications of this search reconfiguration may depend on the contexts in which innovation activities are embedded. We therefore examine two boundary conditions that may shape the relationship between network positions, technological search, and productivity, namely firms' scientific embeddedness and the technological distance involved in their innovation activities.

To empirically investigate these relationships, we construct a time-varying patent citation network based on detailed Chinese patent data, and examine how firms' network centrality influences technological search patterns and productivity. Our analysis jointly observes firms' network positions, technological search patterns, and productivity outcomes. The results show that network centrality is associated with systematic changes in firms' technological search and that these behavioral adjustments mediate the relationship between network position and firm productivity. By linking structural embeddedness with firms' technological search behavior, this study uncovers a behavioral mechanism through which innovation network positions influence productivity. In doing so, it bridges the literature on innovation networks and technological search and provides new insights into the micro-foundations of sustained productivity growth.

The paper is organised as follows. Section~\ref{Sec_Theoretical_framework} presents the theoretical framework. Section~\ref{Sec_Hypotheses} presents the hypotheses. Section~\ref{Sec_Data_method} provides the data and method. Section~\ref{Sec_Results} analyses the results of the study and considers how central firms in innovation networks impact productivity. Section~\ref{Sec_Discussion} discusses the conclusion.

\section{Theoretical framework}
\label{Sec_Theoretical_framework}
Technological innovation is inherently cumulative and path dependent \citep{FJ-Dosi-1982-RP}. Firms' technological progress does not emerge from isolated or random inventive acts. Rather, it evolves through sustained investments and repeated experimentation within specific technological domains \citep{FJ-Cohen-1990-ASQ,FJ-March-1991-OS}. Outcomes of past technological activities, therefore, tend to serve as natural starting points for subsequent innovation efforts \citep{FJ-Cyert-2020-B}. At the same time, the technological knowledge, skills, and experience of R\&D teams and individuals shape how new projects are defined and selected \citep{FJ-Dosi-1982-RP,FJ-Hohberger-2015-RP,FJ-Nelson-1985,FJ-Stuart-1996-SMJ}.

As experience accumulates, individual and team level knowledge becomes institutionalized into firms' cognitive frames and technological search routines. These routines form the basis for how firms interpret new information and evaluate technological opportunities \citep{FJ-Cohen-1990-ASQ,FJ-Gavetti-2012-OS,FJ-Leten-2016-JMS,FJ-Nelson-1985}. Consequently, even when facing identical external technological information, firms differ systematically in their ability to recognize, assess, and exploit technological opportunities.

Firms are generally more capable of identifying, understanding, and recombining knowledge elements that are close to their existing technological base. In contrast, opportunities that lie far from their current knowledge stock are less likely to be considered \citep{FJ-Barreto-2012-OS,FJ-Gregoire-2010-OS,FJ-Gruber-2012-JM}. Under conditions of bounded rationality, managers cannot attend equally to all available technological opportunities \citep{FJ-Hohberger-2015-RP}. As a consequence, technological search becomes systematically localized around familiar domains rather than evenly distributed across the full opportunity space \citep{FJ-Katila-2002-AMJ}. Search routines shaped by historical accumulation constrain firms' search space to areas adjacent to their existing technological resources \citep{FJ-Benner-2012-SMJ,FJ-Christensen-2015-B}. 

Different search directions are associated with distinct learning costs and organizational challenges. Searching along established technological paths allows firms to reuse existing knowledge modules, organizational processes, and evaluation criteria, resulting in more predictable learning processes and lower coordination costs \citep{FJ-Rosenkopf-2001-SMJ}. By contrast, cross-domain entry requires firms to understand unfamiliar technological principles and integrate them with existing capabilities. This increases cognitive burdens and places greater demands on organizational structures and resource allocation \citep{FJ-Leten-2016-JMS,FJ-March-1991-OS,FJ-Xie-2014-SMJ,FJ-Yayavaram-2018-SMJ}. Taken together, the interaction between path dependence and cost considerations implies that firms' choices of technological search direction shape not only the scope of knowledge accumulation, but also their allocation between short-term efficiency gains and long-term adaptive capacity \citep{FJ-March-1991-OS}.

However, firms are not entirely constrained by their past. The very cognitive constraints that arise from path dependence highlight the critical role of external structures in reshaping firms' technological attention and opportunity recognition. Innovation networks, in particular, provide firms with exposure to heterogeneous technological signals and alternative search trajectories. Through repeated interaction and broad network exposure, firms occupying more central positions gain earlier access to diverse technological signals and non-redundant knowledge flows. By altering the informational environment, a firm's position within innovation networks can expand firms' technological attention and reshape their search orientation. These behavioral adjustments are particularly important because they influence how firms allocate innovative efforts and ultimately affect productivity outcomes. Building on this perspective, the following section develops hypotheses on how network centrality shapes technological search behavior and how such behavioral reorientation translates into firm productivity.

\section{Hypotheses}
\label{Sec_Hypotheses}

\subsection{Innovation networks and firm productivity}
A large body of research suggests that firms' embeddedness in interorganizational networks plays a critical role in shaping both their economic and innovative performance \citep{FJ-Kao-2019-IJPE,FJ-Tsai-2001-AMJ,FJ-Uzzi-1996-ASR}. Innovation networks are not merely an external context for firms. Rather, they constitute key channels through which knowledge spillovers and heterogeneous resources circulate. As a result, the performance implications of network structures do not arise simply from an increase in the volume of innovative activity \citep{FJ-Ahuja-2000-ASQ,FJ-Stuart-1996-SMJ}. Instead, innovation networks affect firm performance by shaping how firms access, interpret, and utilize technological knowledge \citep{FJ-Tsai-2001-AMJ}. In particular, these networks provide a structured knowledge environment that governs firms' exposure to technological opportunities \citep{FJ-Lee-2009-T,FJ-Rodriguez-2016-ITEM,FJ-Wang-2014-AMJ}, while also influencing the cognitive and organizational processes through which knowledge is converted into economic value \citep{FJ-Stuart-1996-SMJ,FJ-Tsai-2001-AMJ}. Within such networks, firms occupying more central positions tend to be embedded in influential technological trajectories \citep{FJ-Lyu-2019-TFSC}. They are therefore more likely to engage with technological knowledge that has been widely recognized and validated within the innovation system \citep{FJ-Zaheer-2005-SMJ}. Accordingly, network embeddedness changes the efficiency with which firms learn from prior inventions and recombine existing knowledge components \citep{FJ-Schillebeeckx-2021-JM}. This directly affects how effectively innovative efforts are translated into economic performance. By reshaping these learning processes and search strategies \citep{FJ-Gilsing-2008-RP,FJ-Uzzi-1996-ASR}, network centrality creates systematic differences in how efficiently firms convert technological knowledge into productive output \citep{FJ-Schilling-2007-MS}. As a result, firms holding more central positions in innovation networks are expected to achieve higher levels of productivity \citep{FJ-Giovannetti-2017-IJPE,FJ-Kao-2019-IJPE}.

\begin{itemize}
    \item Hypothesis 1. Firms occupying more central positions in innovation networks exhibit higher productivity.
\end{itemize}

\subsection{The mediating role of technological search behavior in driving productivity}

Open innovation theory posits that firms can no longer rely solely on internal resources to innovate \citep{FJ-Chesbrough-2003-HBP}, but must instead be embedded in broader innovation networks \citep{FJ-Ahuja-2000-ASQ,FJ-Lyu-2019-TFSC,FJ-Zaheer-2005-SMJ} that shape the generation and circulation of technological knowledge across organizational boundaries \citep{FJ-De-2024-JTT,FJ-Wang-2024-TFSC}. Yet, firms face profound resource constraints when determining their technological search strategies. Because expanding exploration and maintaining a diversified technological portfolio both require substantial innovation resources, firms must continuously allocate attention and investment between short-term returns and long-term technological opportunities \citep{FJ-Benner-2003-AMR,FJ-Lavie-2010-AMA,FJ-Stadler-2014-IJMR,FJ-Zhang-2020-TFSC}. In this context, innovation networks serve not merely as channels for acquiring external knowledge, but also as critical informational infrastructures that shape how firms perceive technological opportunities and organize their search processes.

A central position in an innovation network provides firms with privileged access to diverse technological knowledge and a broader vantage point from which to observe technological developments across the network \citep{FJ-Gilsing-2008-RP}. Firms occupying central positions interact with a larger number of partners and are therefore exposed to heterogeneous knowledge originating from multiple technological domains \citep{FJ-Bianchi-2011-T,FJ-Tsai-2001-AMJ}. Such informational advantages expand firms' awareness of emerging technological opportunities and increase the variety of knowledge elements available for potential recombination \citep{FJ-Burt-2000-ROB,FJ-Corsaro-2012-IMM,FJ-Zhang-2020-TFSC}. Through these mechanisms, network centrality reshapes firms' technological search behavior by altering both opportunity recognition processes and the scope of knowledge available for innovation \citep{FJ-Bjork-2009-JPIM,FJ-Lopez-2016-RP}.

From the perspective of search direction, these informational advantages increase firms' likelihood of entering new technological domains \citep{FJ-Gilsing-2008-RP}. Centrally positioned firms can more frequently observe the exploratory activities and technological experimentation of other actors in the network, thereby obtaining valuable external reference points for evaluating unfamiliar technological opportunities \citep{FJ-Ahuja-2000-ASQ,FJ-Lazer-2007-ASQ}. In complex technological landscapes characterized by multiple local optima, firms constrained by entrenched cognitive routines tend to engage in localized search within familiar technological trajectories \citep{FJ-Hohberger-2015-RP,FJ-Stuart-1996-SMJ}. By contrast, the enhanced visibility provided by network centrality reduces the cognitive uncertainty associated with crossing established technological boundaries \citep{FJ-Zhang-2020-TFSC}. As a result, centrally positioned firms are more likely to extend their innovation activities into new technological domains through exploratory search \citep{FJ-Leten-2016-JMS,FJ-Tsai-2001-AMJ,FJ-Yayavaram-2018-SMJ}.

Beyond shaping the direction of technological search, network centrality may also influence the structural configuration of firms' technological portfolios. Exposure to heterogeneous knowledge sources increases the recombination potential of firms' technological knowledge, enabling firms to integrate disparate knowledge components into their innovation activities \citep{FJ-Burt-2000-ROB,FJ-Corsaro-2012-IMM}. Prior research suggests that innovation frequently emerges from the recombination of previously unconnected knowledge elements \citep{FJ-Fleming-2001-MS,FJ-March-1991-OS}. Firms embedded in central network positions therefore possess greater opportunities to access and combine knowledge from different technological domains \citep{FJ-Ahuja-2000-ASQ,FJ-Tsai-2001-AMJ}. Such informational diversity encourages firms to distribute innovative efforts across a broader range of technological fields rather than concentrating on a narrow set of established trajectories \citep{FJ-March-1991-OS}. Consequently, centrally positioned firms are more likely to maintain more diversified technological portfolios.

Based on the above arguments, we propose the following hypotheses:
\begin{itemize}
\item Hypothesis 2a. Firms occupying more central positions in innovation networks are more likely to engage in exploratory technological search by entering new technological domains.

\item Hypothesis 2b. Firms occupying more central positions in innovation networks are more likely to maintain more diversified technological portfolios.
\end{itemize}

Search direction adjustments induced by firms' network positions constitute a critical micro-behavioral foundation for improvements in total factor productivity. By reshaping the structure of technological search, firms can improve the efficiency with which innovation resources are allocated and, over time, develop more competitive productivity trajectories \citep{FJ-Cohen-1990-ASQ,FJ-Fleming-2001-MS,FJ-Katila-2002-AMJ,FJ-March-1991-OS}. In particular, exploratory technological search introduces heterogeneous knowledge combinations that enable firms to reconfigure their production functions \citep{FJ-Levinthal-1993-SMJ}, thereby increasing long-term productivity potential \citep{FJ-Dosi-1982-RP,FJ-Lee-2003-MS} and allowing firms to escape the trap of diminishing marginal returns associated with mature technological domains.

Although exploratory search is often accompanied by substantial adjustment costs in the short run \citep{FJ-Katila-2002-AMJ,FJ-Levinthal-1993-SMJ}, the recombination opportunities and dynamic learning benefits it generates can gradually be transformed into systematic improvements in total factor productivity as new capabilities are integrated and stabilized \citep{FJ-Bloom-2013-Econ}. Conversely, sticking excessively to local, familiar search domains tends to reinforce technological concentration, which can trap firms in persistent efficiency losses. To mitigate such traps and unlock further efficiency gains, firms also benefit from broadening the structure of their technological portfolios. Increasing technological diversity expands the space for knowledge recombination and reduces the risks of technological lock-in \citep{FJ-Katila-2002-AMJ}. While drawing inspiration from familiar knowledge components and refining existing combinations can increase the average success rate of innovation activities and reduce outcome uncertainty \citep{FJ-Fleming-2001-MS}, such self-imposed constraints on the recombination space significantly limit the upside potential for generating high-impact or disruptive inventions \citep{FJ-Zhang-2020-TFSC}.

In the long run, specialization within narrowly defined technological domains may enhance the likelihood of achieving immediate positive returns, but it also increases the risk of technological obsolescence and erodes organizational flexibility in accessing frontier developments \citep{FJ-Sorensen-2000-ASQ}. Accordingly, when network centrality encourages firms to diversify their technological portfolios beyond established technological trajectories, such avoidance of success traps enables more efficient cross-domain allocation of resources, thereby enhancing overall productivity \citep{FJ-Fitzgerald-2021-MS,FJ-Stadler-2014-IJMR}.

\begin{itemize}
    \item Hypothesis 3a. Innovation network centrality enhances firm productivity by inducing firms to engage in exploratory technological search.
    \item Hypothesis 3b. Innovation network centrality enhances firm productivity by increasing the diversity of firms' technological portfolios.
\end{itemize}

\subsection{The moderating role of scientific knowledge sources}

Firms occupying central positions in innovation networks gain access to a wide array of heterogeneous technological information and emerging opportunities through their ties with diverse innovation partners \citep{FJ-Bianchi-2011-T,FJ-Garcia-2024-IJIO,FJ-Lyu-2019-TFSC}. However, merely acquiring information does not necessarily translate into exploratory technological search, which involves entering new technological domains and developing unfamiliar knowledge bases \citep{FJ-Gilsing-2008-RP,FJ-Graf-2022-SBE}. In other words, moving beyond existing technological trajectories requires not only external signals but also the underlying knowledge structures that support the entry into and implementation of new technological paths \citep{FJ-Cohen-1990-ASQ,FJ-Zahringer-2017-ICC}.

In this regard, sources of scientific knowledge from universities and public research institutions can provide critical cognitive and organizational support \citep{FJ-Bellucci-2016-JTT,FJ-Garcia-2024-IJIO}. Prior research suggests that scientific knowledge is typically situated upstream in the technological evolution process and possesses high levels of generality, universality, and cross-domain applicability. As an external input, scientific knowledge can significantly reduce search costs in innovation processes and help firms understand the broader technological landscape at a lower cost \citep{FJ-Fleming-2004-SMJ,FJ-Zahringer-2017-ICC}. Compared with applied knowledge that primarily serves to optimize existing technological paths, scientific knowledge provides firms with a more universal knowledge base, enabling more structured search in complex problem contexts \citep{FJ-Klevorick-1995-RP}. It also makes it easier for firms to recombine dispersed informational cues into new technological directions, thereby facilitating higher levels of novelty in innovation \citep{FJ-Kohler-2012-RP,FJ-Slavova-2021-T}. As \cite{FJ-Fleming-2004-SMJ} argued, science can tell inventors how to avoid wasting effort, thereby enhancing the efficiency and directionality of exploratory search.

For centrally positioned firms, these benefits of scientific knowledge serve as a critical complement to network advantages. On the one hand, network centrality increases the likelihood of encountering new technological cues. On the other hand, a higher level of scientific knowledge provides the necessary foundational support for entering unfamiliar technological domains, reducing the cognitive barriers and organizational uncertainty associated with exploratory innovation \citep{FJ-Klevorick-1995-RP,FJ-Slavova-2021-T}. Under such conditions, central firms are more likely to translate the heterogeneous information acquired through their networks into exploratory R\&D activities with breakthrough potential \citep{FJ-Lian-2025-T}. Conversely, when firms rely less on scientific knowledge, even a central network position may not enable effective cross-domain exploration; instead, such firms are more likely to pursue incremental improvements along existing technological paths, thereby reinforcing path dependence \citep{FJ-March-1991-OS}.
\begin{itemize}
    \item Hypothesis 4. Scientific knowledge sources positively moderate the relationship between innovation network centrality and exploratory technological search.
\end{itemize}

\subsection{The moderating role of technological distance}

Technological distance involves two countervailing forces that shape innovation performance. On the one hand, more distant technological domains are characterized by greater uncertainty and lower visibility, which tends to amplify the informational benefits of network centrality. On the other hand, greater technological distance raises the costs of absorbing and integrating new knowledge, making it harder to translate exploratory and diversified search into productivity gains.

Exploratory technological search allows firms to expand their knowledge base and enter new technological domains \citep{FJ-Rosenkopf-2001-SMJ,FJ-Zhu-2021-T}. By extending their search beyond existing technological trajectories, firms may discover alternative technological solutions and introduce new knowledge elements that stimulate novel combinations, thereby contributing to improvements in firm productivity. However, the productivity gains from exploratory search are constrained by the technological distance between newly explored domains and the firm's existing knowledge base.

Technological distance reflects the degree of dissimilarity between unfamiliar technologies and a firm's established technological capabilities \citep{FJ-Gilsing-2008-RP}. When exploratory activities target technologically proximate domains, firms can more easily recombine new knowledge with their existing capabilities and integrate it into ongoing technological and production processes. In contrast, when exploration extends into technologically distant domains, the challenges associated with knowledge integration increase substantially \citep{FJ-Benner-2003-AMR,FJ-Quintana-2008-RP,FJ-Vedel-2021-RP}. Distant technologies often rely on different scientific principles, engineering architectures, and production logics, making it more difficult for firms to effectively integrate new knowledge with their existing technological foundations \citep{FJ-Nelson-1985,FJ-Rosenkopf-2001-SMJ}. As a result, although exploratory search may still generate new knowledge, the likelihood that such knowledge can be successfully transformed into productivity improvements decreases as technological distance increases. A similar logic applies to technological diversity: greater distance reduces the complementarity between knowledge components and raises the costs of cross-domain recombination.

Meanwhile, technological distance also moderates the value of network centrality in a distinct way. Although distant technological domains introduce substantial uncertainty regarding the relevance and applicability of unfamiliar knowledge, this uncertainty also makes the structural advantages of a central network position particularly valuable \citep{FJ-Lyu-2019-TFSC,FJ-Wen-2021-T}. Under conditions of high technological distance, firms struggle to judge which unfamiliar knowledge can be feasibly integrated into their existing systems. In this setting, centrally positioned firms benefit from broader information flows and superior visibility of cross-network technological developments \citep{FJ-Gilsing-2008-RP,FJ-Tsai-2001-AMJ}.These informational advantages allow them to better evaluate distant opportunities and identify knowledge that remains relevant to their technological foundations \citep{FJ-Fleming-2004-SMJ,FJ-Nelson-1985}. As a result, the productivity-enhancing effect of network centrality becomes stronger when technological distance is high. In short, centrality helps firms mitigate the frictions associated with distant technological search.

\begin{itemize}
\item Hypothesis 5a. Technological distance negatively moderates the relationship between technological exploration, technological diversity and firm productivity.
\end{itemize}
\begin{itemize}
\item Hypothesis 5b. Technological distance positively moderates the relationship between innovation network centrality and firm productivity.
\end{itemize}

\begin{figure}[htb!]
\centering
\includegraphics[width=1\linewidth]{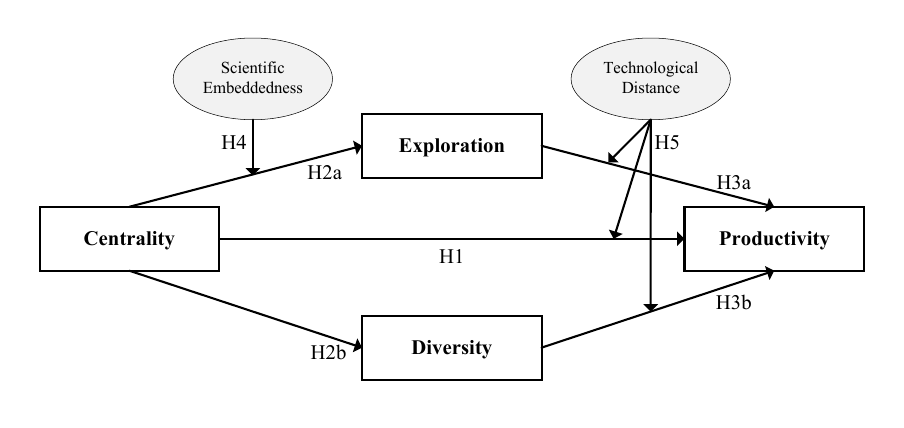}
\caption{The integrated theoretical framework with hypotheses.}
\label{Fig_framework}
\end{figure}

Fig.~\ref{Fig_framework} presents the integrated theoretical framework of this study, summarizing the hypothesized relationships.

\section{Data and method}
\label{Sec_Data_method}

\subsection{Data}

Our analysis draws on two primary data sources: patent data from the China National Intellectual Property Administration\footnote{CNIPA, \url{https://www.cnipa.gov.cn}.}, and firm-level information from the China Stock Market and Accounting Research Database\footnote{CSMAR, \url{https://data.csmar.com}.}.

Firstly, patent data are obtained from the CNIPA and cover the period from 1985 to 2023. The dataset includes comprehensive information on all patent applications filed in China, such as patent publication announcement numbers, application numbers, application years, patent types, cited patents, applicant names, IPC classification codes and citation information and so on. Using these data, we construct China's patent innovation network at the applicant level. Each node represents a patent applicant, and directed edges are defined by patent citation relationships. Importantly, the network incorporates all types of innovation entities, including publicly listed firms, non-listed firms, universities, and public research institutions. Then we use the network to compute network indicators, including PageRank, which capture each applicant’s positions in the national innovation system.

Second, financial information for publicly listed companies is obtained from CSMAR database. The database provides detailed data on firms' financial statements, operating characteristics, ownership structure, R\&D activities, and other key control variables. We identify publicly listed firms from the patent applicant information and merge the patent-based network indicators with firm-level data using firm identifiers and year information. Firms with fewer than three observations are excluded, yielding a final unbalanced panel spanning 2010 to 2023.

\subsection{Variables}
\subsubsection{Dependent variable}
We measure our dependent variable using firm-level total factor productivity (TFP), which captures firms' efficiency in transforming inputs into output. Estimating TFP is challenging because unobserved productivity shocks may influence firms' input choices, leading to simultaneity bias. To address this issue, we follow \cite{FJ-Olley-1992-NBER}, which uses firms' investment decisions as a proxy for unobserved productivity, allowing consistent estimation of input elasticities.
Specifically, we assume a Cobb-Douglas production function in which firm output is produced using capital and labor as inputs. Capital is measured by firms' net fixed assets, and labor is proxied by the number of employees. Following standard practice, investment is derived from the evolution of capital according to the law of motion $K_{t}=(1-\delta)K_{t-1}+I$, where $\delta$ is the depreciation rate. In practice, we measure investment as the net increase in fixed assets plus depreciation.
Based on the estimated production function, we construct TFP as the residual component of output that cannot be explained by observable inputs. This measure captures firms' efficiency in transforming inputs into output and serves as the main outcome variable in our empirical analysis. For robustness checks, we consider alternative measures of TFP estimated using different production function approaches.

\subsubsection{Independent variable}
Our key independent variable is firms' position in the patent innovation network, denoted as $Centrality$. Following \cite{FJ-Brin-1998-CNaIS}, we measure firms' network centrality using PageRank centrality, which captures the recursive importance of nodes based on incoming citation links. The $Centrality$ value is calculated as follows:
\begin{equation}
{Centrality}_{i,t}= \frac{1-d}{N_{t}}+ d\sum_{j \in M(i)}\frac{{Centrality}_{j,t}}{C_{j,t}},
\end{equation}
where $d$ is the damping factor. Following prior studies and the original PageRank algorithm, the damping factor $d$ is set to 0.85 \citep{FJ-Brin-1998-CNaIS}. $N_{t}$ denotes the total number of applicants in the network in year $t$, and $C_{j,t}$ is the number of outgoing citation links from node $j$. PageRank values are computed iteratively until convergence.

To construct a time-varying measure of firms' network positions, we build an annual patent citation subnetwork using a rolling five-year window. Specifically, for each year $t$, we retain all invention patents applied for during the period $[t-4, t]$ and the associated citation relationships. We then construct an applicant-level innovation network in which nodes represent patent applicants and directed links capture citation relationships aggregated from patent-level citations. Citations are included if the citing patent falls within the focal rolling window, ensuring temporal consistency in knowledge flows. Patent-level information is subsequently aggregated to the applicant level, and each year thus corresponds to a distinct innovation network on which the PageRank centrality of each applicant is computed.

A firm's centrality reflects its structural importance within the technological knowledge flow, capturing the extent to which the firm occupies influential and well-connected positions in the innovation network. Because PageRank recursively weights citations by the centrality of the citing nodes, a higher value of $Centrality$ indicates that a firm's technologies are recognized by influential innovators and embedded in prominent technological trajectories. Firms occupying such positions are closely connected to leading innovators whose subsequent developments frequently build upon the focal firm's knowledge base. As a result, these firms are more likely to observe the evolution of emerging technologies and shifts in innovation directions across the network. In addition, $Centrality$ is standardized to facilitate interpretation and reduce the influence of outliers.

\subsubsection{Mechanism variables}
To empirically examine firms' technological search behavior, we rely on observable patterns in the structure of firms' patent portfolios. Rather than directly inferring firms' search intentions, our approach focuses on realized innovation activities reflected in how firms expand and recombine knowledge across technological domains. Patent portfolios provide a useful representation of firms' technological knowledge bases and therefore allow us to observe how firms adjust their innovation activities over time. Prior research suggests that firms' technological search can vary along multiple dimensions. In particular, firms may adjust the direction of search by entering new technological domains, and they may also adjust the breadth of search by distributing innovation activities across a wider range of knowledge areas. These dimensions capture different aspects of how firms reconfigure their technological knowledge bases when responding to new opportunities. Accordingly, we operationalize technological search behavior using two complementary indicators. First, technological exploration captures firms' entry into technological domains that were absent from their recent innovation portfolios, reflecting a shift toward new areas of knowledge. Second, technological diversity measures the breadth of technological fields represented in firms' patent portfolios, indicating the extent to which innovation activities are distributed across heterogeneous knowledge domains. Together, these two indicators capture how firms reconfigure their technological search through both exploratory entry and broader technological scope.

\paragraph{Technological Diversity}
Technological diversity captures the extent to which a firm's innovative activities are distributed across technological domains, thereby reflecting the structural orientation of technological search. In the empirical analysis, $Diversity$ is operationalized by the Herfindahl-Hirschman Index ($HHI$), constructed from the distribution of a firm's patents across IPC classes. Formally, for firm $i$ in year $t$:

\begin{equation}
Diversity_{i,t} =1-\sum_{k} s_{i,k,t}^2,
\end{equation}
where $s_{i,k,t}$ denotes the share of patents held by firm $i$ in IPC class $k$ in year $t$. The index ranges from 0 to 1, with higher values indicating that a firm's innovative activities are distributed across a broader set of technological domains. Lower values reflect technological concentration within a limited number of domains, suggesting a stronger reliance on familiar technological trajectories.

\paragraph{Technological Exploration}
Technological exploration captures firms' expansion into unfamiliar knowledge domains and reflects exploratory search orientation. Inspired by prior studies that identify exploratory technological activities based on entries into previously unused technological components \citep{FJ-Guan-2016-RP,FJ-Yayavaram-2015-SMJ,FJ-Zakaryan-2023-T,FJ-Wen-2021-T}, we construct a domain-based measure of exploration breadth. Specifically, $Exploration$ is defined as the number of IPC four-digit classes newly entered by a firm in year $t$ that were absent from its technological portfolio during the preceding five years. This five-year historical window is consistent with the rolling innovation window used in constructing the patent citation network, thereby maintaining temporal alignment across key variables. By capturing entries into previously unused technological classes, $Exploration$ reflects firms' outward-oriented technological search beyond existing knowledge trajectories and indicates stronger exploratory innovation behavior.

\subsubsection{Moderating Variables}
To identify the boundary conditions of search transformation and its efficiency outcomes, we construct two contextual variables. 

\paragraph{Scientific Embeddedness}
Scientific embeddedness reflects the extent to which a firm relies on scientific knowledge sources in its innovation activities. We measure $SciEmbed$ as the proportion of a firm's backward patent citations that are directed toward universities and public research institutes within the patent citation network. Formally, for firm $i$ in year $t$:
\begin{equation}
SciEmbed_{i,t} = \frac{\text{Citations to universities and public research institutes}_{i,t}}{\text{Total backward citations}_{i,t}},
\end{equation}
A higher value of $SciEmbed$ indicates a stronger embedding in scientific knowledge sources, which provides a more general and principle-based knowledge base. Conceptually, $SciEmbed$ functions as a \emph{navigation mechanism} that enhances firms' ability to identify, interpret, and act upon frontier technological signals encountered through their network positions.

\paragraph{Technological Distance}
Technological distance captures the degree of dissimilarity between a firm's newly explored technological domains and its historical technological knowledge base. We operationalize technological distance using the cosine distance between IPC-based technology vectors. Specifically, for firm $i$ in year $t$, technological distance is defined as:
\begin{equation}
TechDist_{i,t} = 1 - \frac{\mathbf{V}_{i,t} \cdot \mathbf{V}_{i,\text{history}}}{\|\mathbf{V}_{i,t}\| \, \|\mathbf{V}_{i,\text{history}}\|},
\end{equation}
where $\mathbf{V}_{i,t}$ denotes the IPC-based vector representing the technological composition of the firm's newly entered domains in year $t$, and $\mathbf{V}_{i,\text{history}}$ represents the IPC-based vector capturing the firm's accumulated technological portfolio prior to year $t$.
A higher value of $TechDist$ indicates greater technological dissimilarity, implying higher recombination complexity and uncertainty associated with cross-domain technological search.

\subsubsection{Control variables}
Following previous studies, we include control variables that capture a firm's intrinsic properties, financial status, and R\&D activities. For intrinsic properties, we control for a firm's age (\textit{Age}), labor input (\textit{Labor}), and capital (\textit{Capital}). \textit{Age} is measured as the difference between the year of firm establishment and the focal year, reflecting firm maturity and accumulated experience in innovation. \textit{Labor} and \textit{Capital} are measured as the natural logarithm of the number of employees and the book value of fixed assets, respectively, capturing the scale of human and physical capital available for production and innovation activities. Together, these variables represent the primary inputs of the production process. For a firm's financial status, we include return on assets (\textit{ROA}) and debt-to-assets ratio (\textit{Lev}). \textit{ROA} reflects firm profitability, while \textit{Lev} indicates financial pressure, which may influence risk-taking and innovation behavior. For R\&D activities, we include R\&D intensity (\textit{RD}), defined as the ratio of R\&D expenditures to operating revenue. \textit{RD} captures the firm's proactive investment in knowledge creation and its potential effect on productivity outcomes. Notably, we do not include a separate firm size variable, as \textit{Labor} and \textit{Capital} together adequately capture firm scale while maintaining consistency with the production function framework used to estimate productivity.

\subsection{Methodology}
To examine the relationship between firms' positions in the patent innovation network and productivity, we estimate panel regressions of the following baseline specification:
\begin{equation}
TFP_{i,t}= \beta_{0}+ \beta_{1}Centrality_{i,t}+ \mathbf{V}_{i,t}\boldsymbol{\varphi}
+ \lambda_t + \delta_j + \gamma_c + \varepsilon_{i,t},
\end{equation}
where $TFP_{i,t}$ denotes the productivity of firm $i$ in year $t$, and $Centrality_{i,t}$ measures the firm's centrality in the innovation network. $\mathbf{V}_{i,t}$ is a vector of control variables discussed in previous section, capturing firms' intrinsic characteristics, financial conditions, and R\&D activities. $\lambda_t$, $\delta_j$, and $\gamma_c$ denote year, industry, and city fixed effects, respectively. We include year fixed effects to control for common macroeconomic shocks and time-varying institutional factors. In addition, we control for industry and city fixed effects to account for persistent differences in technological opportunities across industries and regional innovation environments across locations. Standard errors are clustered at the firm level to allow for arbitrary serial correlation within firms over time. We do not include firm fixed effects in the baseline specification, as firms' positions in the patent innovation network are highly persistent and largely reflect long-term structural characteristics. Including firm fixed effects would absorb much of the relevant variation in network centrality. We nevertheless examine specifications with firm fixed effects as a robustness check.

\section{Results}
\label{Sec_Results}

\subsection{The impact of firm's innovation centrality on total factor productivity}

Table~\ref{Table_BaselineRegression} reports the baseline results on the relationship between firms' innovation centrality and total factor productivity. Model (1) includes only control variables, while Model (2) further incorporates the core explanatory variable, innovation centrality, measured as $Centrality$ in the innovation network. As shown in Model (2), PR is positively and significantly associated with productivity ($\beta=0.177$, $p<0.01$), indicating that firms occupying more central positions in the innovation network exhibit higher productivity. This finding provides strong support for Hypothesis 1.
 
\begin{table}[htb!]
\centering
\renewcommand\arraystretch{0.8}
\renewcommand{\tabcolsep}{12.2mm}
\caption{Baseline results}
\label{Table_BaselineRegression}
\begin{tabular}{lcc}
\toprule
 & \multicolumn{2}{c}{Dependent variable: Total Factor Productivity ($TFP$)} \\
 \cmidrule(lr){2-3}
 & (1) & (2) \\
\midrule
$Centrality$        &               & 0.177$^{***}$ \\
          &               & (0.018)       \\

$Age$       & 0.064         & 0.046         \\
          & (0.063)       & (0.060)       \\

$Labor$     & 0.096$^{***}$ & 0.046$^{*}$   \\
          & (0.025)       & (0.024)       \\

$Capital$   & 0.098$^{***}$ & 0.071$^{***}$ \\
          & (0.022)       & (0.020)       \\

$RD$        & $-$2.863$^{***}$ & $-$3.166$^{***}$ \\
          & (0.417)       & (0.406)       \\

$ROA$       & 0.020$^{*}$   & 0.019$^{*}$   \\
          & (0.010)       & (0.010)       \\

$Lev$       & 0.530$^{***}$ & 0.559$^{***}$ \\
          & (0.107)       & (0.103)       \\

Constant     & 11.043$^{***}$ & 12.077$^{***}$ \\
          & (0.378)       & (0.372)       \\

\midrule
Year FE   & YES & YES \\
Ind FE    & YES & YES \\
City FE   & YES & YES \\

Observations & 5,121 & 5,121 \\
$R^2$-adj & 0.583 & 0.609 \\
\bottomrule
\end{tabular}

\begin{flushleft}
\footnotesize
\textit{Notes}: The dependent variable is firm-level total factor productivity, estimated using the OP method \citep{FJ-Olley-1992-NBER}. Standard errors clustered at the firm level are reported in parentheses. 
$^{***}$, $^{**}$, and $^{*}$ denote significance at the 1\%, 5\%, and 10\% levels, respectively.
\end{flushleft}
\end{table}

\subsection{Robustness check}

To ensure that our main results are not driven by specific measurement choices or model specifications, we conduct several robustness checks. Table~\ref{Table_Robustness} reports the results.

First, we replace the baseline $TFP$ measure estimated by the OP approach with alternative $TFP$ measures. Models (1) and (2) use $TFP$ estimated by OLS and the LP method \citep{FJ-Levinsohn-2003-RES}, respectively. The coefficient on $Centrality$ remains positive and statistically significant in both specifications, with similar magnitudes, suggesting that our findings are not sensitive to the choice of productivity estimation method.

Second, we examine whether the results are robust to alternative measures of firms' positions in the patent citation network. Models (3)-(5) replace PageRank centrality with eigenvector centrality ($E\_Centrality$), out-degree centrality ($OD\_Centrality$), and out-strength centrality ($OS\_Centrality$), respectively. These alternative indicators capture different dimensions of network embeddedness, such as influence, connectivity, and the intensity of knowledge outflows. Across all specifications, the estimated coefficients on network position remain positive and highly significant, indicating that the productivity enhancing effect is a general property of firms' network positions rather than being driven by a specific centrality measure.

Finally, Model (6) adopts a more stringent fixed effects structure by replacing industry and city fixed effects with firm fixed effects while keeping year fixed effects. This specification absorbs time invariant firm level heterogeneity, including persistent differences in managerial quality, technological capability, and organizational structure. The coefficient on $Centrality$ remains statistically significant, further supporting the robustness of the baseline results against omitted time-invariant firm characteristics.

Overall, these robustness checks confirm that the positive relationship between firms' innovation centrality and productivity is stable across alternative variable definitions and more demanding fixed-effects specifications, reinforcing the evidence for Hypothesis~1.

\begin{table}[htb!]
\centering
\renewcommand\arraystretch{0.8}
\setlength{\tabcolsep}{2.2pt}
\caption{Robustness check}
\label{Table_Robustness}
\begin{tabular*}{\linewidth}{@{\extracolsep{\fill}}lcccccc}
\toprule
 & \multicolumn{2}{c}{Alternative TFP measures}
 & \multicolumn{3}{c}{Alternative centrality measures}
 & Alternative FE \\
\cmidrule(lr){2-3} \cmidrule(lr){4-6} \cmidrule(lr){7-7}
 & (1) & (2) & (3) & (4) & (5) & (6) \\
\cmidrule(lr){2-3} \cmidrule(lr){4-6} \cmidrule(lr){7-7}
 & $TFP$(OLS) & $TFP$(LP) & $TFP$ & $TFP$ & $TFP$ & $TFP$ \\
\midrule
$Centrality$ & 0.180$^{***}$ & 0.152$^{***}$ &  &  &  & 0.061$^{**}$ \\
   & (0.018) & (0.021) &  &  &  & (0.024) \\

$E\_Centrality$ &  &  & 0.173$^{***}$ &  &  &  \\
   &  &  & (0.017) &  &  &  \\

$OD\_Centrality$ &  &  &  & 0.177$^{***}$ &  &  \\
        &  &  &  & (0.016) &  &  \\

$OS\_Centrality$ &  &  &  &  & 0.172$^{***}$ &  \\
        &  &  &  &  & (0.018) &  \\

$Age$ & 0.043 & 0.069 & 0.047 & 0.089 & 0.096 & 0.131 \\
    & (0.060) & (0.062) & (0.060) & (0.060) & (0.060) & (0.192) \\

$Labor$ & 0.008 & 0.504$^{***}$ & 0.042$^{*}$ & 0.034 & 0.044$^{*}$ & 0.053 \\
      & (0.024) & (0.025) & (0.024) & (0.024) & (0.024) & (0.061) \\

$Capital$ & $-$0.056$^{***}$ & $-$0.014 & 0.073$^{***}$ & 0.072$^{***}$ & 0.076$^{***}$ & $-$0.030 \\
        & (0.020) & (0.021) & (0.020) & (0.020) & (0.020) & (0.024) \\

$RD$ & $-$3.139$^{***}$ & $-$3.164$^{***}$ & $-$3.237$^{***}$ & $-$3.325$^{***}$ & $-$3.228$^{***}$ & $-$4.710$^{***}$ \\
   & (0.399) & (0.410) & (0.399) & (0.398) & (0.399) & (0.342) \\

$ROA$ & 0.019$^{*}$ & 0.019$^{*}$ & 0.019$^{*}$ & 0.019$^{*}$ & 0.019$^{*}$ & 0.009$^{***}$ \\
    & (0.010) & (0.010) & (0.010) & (0.010) & (0.010) & (0.002) \\

$Lev$ & 0.548$^{***}$ & 0.585$^{***}$ & 0.561$^{***}$ & 0.544$^{***}$ & 0.553$^{***}$ & 0.177$^{**}$ \\
    & (0.102) & (0.104) & (0.103) & (0.101) & (0.102) & (0.075) \\

Constant & 11.932$^{***}$ & 11.892$^{***}$ & 12.078$^{***}$ & 12.029$^{***}$ & 11.848$^{***}$ & 14.107$^{***}$ \\
      & (0.364) & (0.384) & (0.367) & (0.364) & (0.362) & (0.553) \\

\midrule
Year FE & YES & YES & YES & YES & YES & YES \\
Ind  FE & YES & YES & YES & YES & YES &  \\
City FE & YES & YES & YES & YES & YES &  \\
Firm FE &  &  &  &  &  & YES \\
\midrule
Observations & 5,121 & 5,121 & 5,121 & 5,121 & 5,121 & 5,122 \\
$R^2$-adj & 0.480 & 0.781 & 0.611 & 0.614 & 0.609 & 0.938 \\
\bottomrule
\end{tabular*}

\begin{flushleft}
\footnotesize
\textit{Notes}: Standard errors clustered at the firm level are reported in parentheses. 
$^{***}$, $^{**}$, and $^{*}$ denote significance at the 1\%, 5\%, and 10\% levels, respectively.
\end{flushleft}
\end{table}

\subsection{Endogeneity Test}

Table~\ref{Table_EndogeneityTests} reports a series of additional analyses designed to address potential endogeneity concerns in the relationship between firms' positions in the innovation network and their productivity.

A first concern relates to simultaneity and reverse causality, whereby contemporaneous productivity shocks may mechanically affect firms' observed network positions. To mitigate this concern, we exploit a lag structure and replace the contemporaneous measure of innovation centrality with its one year lag. Since current productivity cannot influence past network positions, this specification provides a simple and intuitive test against simultaneity bias. The estimated coefficient on lagged innovation centrality remains positive and highly significant ($\beta=0.162$, $p<0.01$), suggesting that the baseline results are unlikely to be driven by contemporaneous reverse causality.

A related concern is that unobserved firm characteristics may simultaneously drive both innovation centrality and productivity, leading to omitted variable bias. To address this issue, we adopt an instrumental variable strategy based on industry level variation in innovation centrality \citep{FJ-Bascle-2008-SO}. Specifically, for each firm $i$ in industry $j$ and year $t$, we construct an instrument by computing the average centrality of all other firms in the same industry-year cell (excluding firm $i$). This industry average centrality captures exogenous shifts in the technological environment and network structure that affect a firm's position through peers' citation behavior, but is unlikely to directly affect the focal firm's productivity conditional on firm controls and fixed effects. Column (2) reports IV estimates using this industry average centrality instrument, and the second stage coefficient on $Centrality$ remains positive and significant ($\beta=0.212$, $p<0.01$). The first stage coefficient on the instrument is positive and statistically significant, and first stage diagnostics indicate that the instrument is sufficiently strong, with Kleibergen-Paap F-statistics well above conventional thresholds and the Kleibergen-Paap LM test rejecting the null of under-identification.

In addition, Column (3) presents estimates using a Bartik instrument constructed from industry-level patenting growth interacted with firm-specific predetermined exposure, following \cite{FJ-Goldsmith-2020-AER}. The second stage coefficient remains positive and statistically significant ($\beta=0.297$, $p<0.01$), and diagnostic tests again confirm instrument relevance and validity. It should be noted that the sample sizes differ across specifications due to data availability for the IV instruments. Nonetheless, the results remain qualitatively consistent across different samples.

Taken together, the evidence from lagged specifications and IV estimations provides consistent support for a causal interpretation of the baseline results. Firms occupying more central positions in the innovation network experience significantly higher productivity, and this effect is robust to a wide range of identification strategies designed to address potential endogeneity.

\begin{table}[htb!]
\centering
\renewcommand\arraystretch{0.8}
\setlength{\tabcolsep}{2.2pt}
\caption{Endogeneity tests: Lag structure and IV estimations}
\label{Table_EndogeneityTests}
\begin{tabular*}{\linewidth}{@{\extracolsep{\fill}}lccc}
\toprule
 & \multicolumn{3}{c}{Dependent variable: Total Factor Productivity ($TFP$)} \\
 \cmidrule(lr){2-4}
 & (1) & (2) & (3) \\
\cmidrule(lr){2-2} \cmidrule(lr){3-3} \cmidrule(lr){4-4}
 & Lagged Centrality & Industry IV & Bartik IV \\
\midrule
$Centrality$ &  & 0.212$^{***}$ & 0.297$^{***}$ \\
   &  & (0.045) & (0.054) \\
$L.Centrality$ & 0.162$^{***}$ &  &  \\
     & (0.019) &  &  \\
$Age$ & 0.065 & 0.044 & 0.023 \\
    & (0.067) & (0.061) & (0.065) \\
$Labor$ & 0.049$^{*}$ & 0.035 & 0.005 \\
      & (0.026) & (0.027) & (0.027) \\
$Capital$ & 0.080$^{***}$ & 0.067$^{***}$ & 0.050$^{**}$ \\
        & (0.022) & (0.021) & (0.022) \\
$RD$ & $-$3.086$^{***}$ & $-$3.212$^{***}$ & $-$3.238$^{***}$ \\
   & (0.441) & (0.420) & (0.415) \\
$ROA$ & 0.016 & 0.019$^{*}$ & 0.017$^{**}$ \\
    & (0.010) & (0.010) & (0.009) \\
$Lev$ & 0.558$^{***}$ & 0.563$^{***}$ & 0.591$^{***}$ \\
    & (0.111) & (0.102) & (0.106) \\
Constant & 11.843$^{***}$ &  &  \\
       & (0.410) &  &  \\
\midrule
Year FE & YES & YES & YES \\
Industry FE & YES & YES & YES \\
City FE & YES & YES & YES \\
KP F stat &  & 23.342 & 46.327 \\
KP LM $p$-val &  & 0.000 & 0.000 \\
Observations & 3,484 & 5,052 & 4,625 \\
\bottomrule
\end{tabular*}

\begin{flushleft}
\footnotesize
\textit{Notes}: Column (1) reports lagged $Centrality$ estimation; columns (2)-(3) report IV estimates using industry average centrality and Bartik instruments, respectively. Standard errors clustered at the firm level are in parentheses; $^{***}$, $^{**}$, $^{*}$ denote significance at the 1\%, 5\%, and 10\% levels, respectively.
\end{flushleft}
\end{table}

\subsection{Mechanism}
Firms positioned at the center of innovation networks are exposed to diverse technological signals and heterogeneous knowledge flows. Beyond mere access to information, network centrality can reshape how firms search for and recombine knowledge along two dimensions: direction, reflected in exploratory entry into new technological domains, and structure, reflected in the diversification of technological portfolios. Whether central firms can translate adjustments in innovation direction and portfolio structure into productivity gains depends on the mechanisms through which network centrality operates.

Table~\ref{Table_Mechanism} investigates the directional mechanism through which innovation centrality affects productivity. Model (1) shows that centrality is positively and significantly associated with technological diversity ($\beta = 0.006$, $p < 0.01$), suggesting that central firms broaden the structural scope of their technological portfolios. In Model (2), after incorporating diversity into the productivity regression, technological diversity is positively and significantly associated with productivity ($\beta = 0.709$, $p < 0.01$). Model (3) examines the directional dimension of innovation and shows that centrality has a positive and significant effect on exploratory entry into new technological domains ($\beta = 0.986$, $p < 0.01$). Model (4) further indicates that technological exploration is positively associated with productivity ($\beta = 0.014$, $p < 0.01$). Together, these results provide consistent support for the dual mechanisms through which centrality influences productivity: central firms not only diversify their technological portfolios but also explore new technological directions, each of which contributes to productivity gains.

Model (5) includes both technological diversity and exploration simultaneously to examine whether these mechanisms jointly mediate the effect of centrality on productivity. The coefficient of centrality remains positive and significant but decreases slightly compared with Models (2) and (4), suggesting that diversity and exploration jointly account for part of the productivity advantage of central firms. This evidence supports a partial mediation interpretation, indicating that innovation centrality improves productivity by encouraging both shifts in technological direction and expansions in technological structure, while other complementary channels may also contribute.

\begin{table}[htb!]
\centering
\renewcommand\arraystretch{0.8}
\setlength{\tabcolsep}{2pt}
\caption{Evidence on innovation direction as a mechanism}
\label{Table_Mechanism}
\begin{tabular*}{\linewidth}{@{\extracolsep{\fill}}lccccc}
\toprule
 & (1) & (2) & (3) & (4) & (5) \\
\cmidrule(lr){2-6}
 & $Diversity$ & $TFP$ & $Exploration$ & $TFP$ & $TFP$ \\
\midrule
$Centrality$ 
& 0.006$^{***}$ & 0.172$^{***}$ & 0.986$^{***}$ & 0.162$^{***}$ & 0.161$^{***}$ \\
& (0.002) & (0.018) & (0.147) & (0.018) & (0.018) \\
$Diversity$ 
&  & 0.709$^{***}$ &  &  & 0.531$^{**}$ \\
&  & (0.228) &  &  & (0.231) \\
$Exploration$ 
&  &  &  & 0.014$^{***}$ & 0.012$^{***}$ \\
&  &  &  & (0.003) & (0.003) \\
$Age$ 
& 0.014$^{**}$ & 0.037 & $-$0.862$^{***}$ & 0.059 & 0.049 \\
& (0.007) & (0.061) & (0.292) & (0.061) & (0.061) \\
$Labor$ 
& 0.008$^{***}$ & 0.040 & 0.722$^{***}$ & 0.036 & 0.033 \\
& (0.003) & (0.024) & (0.106) & (0.024) & (0.024) \\
$Capital$ 
& 0.007$^{***}$ & 0.066$^{***}$ & 0.358$^{***}$ & 0.066$^{***}$ & 0.063$^{***}$ \\
& (0.002) & (0.020) & (0.084) & (0.020) & (0.020) \\
$RD$ 
& 0.117$^{***}$ & $-$3.249$^{***}$ & 1.574 & $-$3.189$^{***}$ & $-$3.247$^{***}$ \\
& (0.037) & (0.402) & (1.612) & (0.405) & (0.402) \\
$ROA$ 
& $-$0.000 & 0.019$^{*}$ & 0.070$^{**}$ & 0.018$^{*}$ & 0.018$^{*}$ \\
& (0.000) & (0.010) & (0.028) & (0.009) & (0.010) \\
$Lev$ 
& 0.044$^{***}$ & 0.528$^{***}$ & 1.010$^{**}$ & 0.544$^{***}$ & 0.524$^{***}$ \\
& (0.012) & (0.104) & (0.455) & (0.101) & (0.103) \\
Constant 
& 0.622$^{***}$ & 11.636$^{***}$ & $-$6.647$^{***}$ & 12.173$^{***}$ & 11.825$^{***}$ \\
& (0.040) & (0.394) & (1.577) & (0.370) & (0.397) \\
\midrule
Year FE  & YES & YES & YES & YES & YES \\
Ind FE   & YES & YES & YES & YES & YES \\
City FE  & YES & YES & YES & YES & YES \\
\midrule
Observations & 5,121 & 5,121 & 5,121 & 5,121 & 5,121 \\
$R^2$-adj & 0.396 & 0.612 & 0.398 & 0.613 & 0.614 \\
\bottomrule
\end{tabular*}

\begin{flushleft}
\footnotesize
\textit{Notes}: This table examines whether innovation network position affects firm productivity through two mechanisms: ($Diversity$) and $Exploration$. Standard errors clustered at the firm level are reported in parentheses. $^{***}$, $^{**}$, and $^{*}$ denote significance at the 1\%, 5\%, and 10\% levels, respectively.
\end{flushleft}
\end{table}

\begin{figure}[htb!]
\centering
\includegraphics[width=1\linewidth]{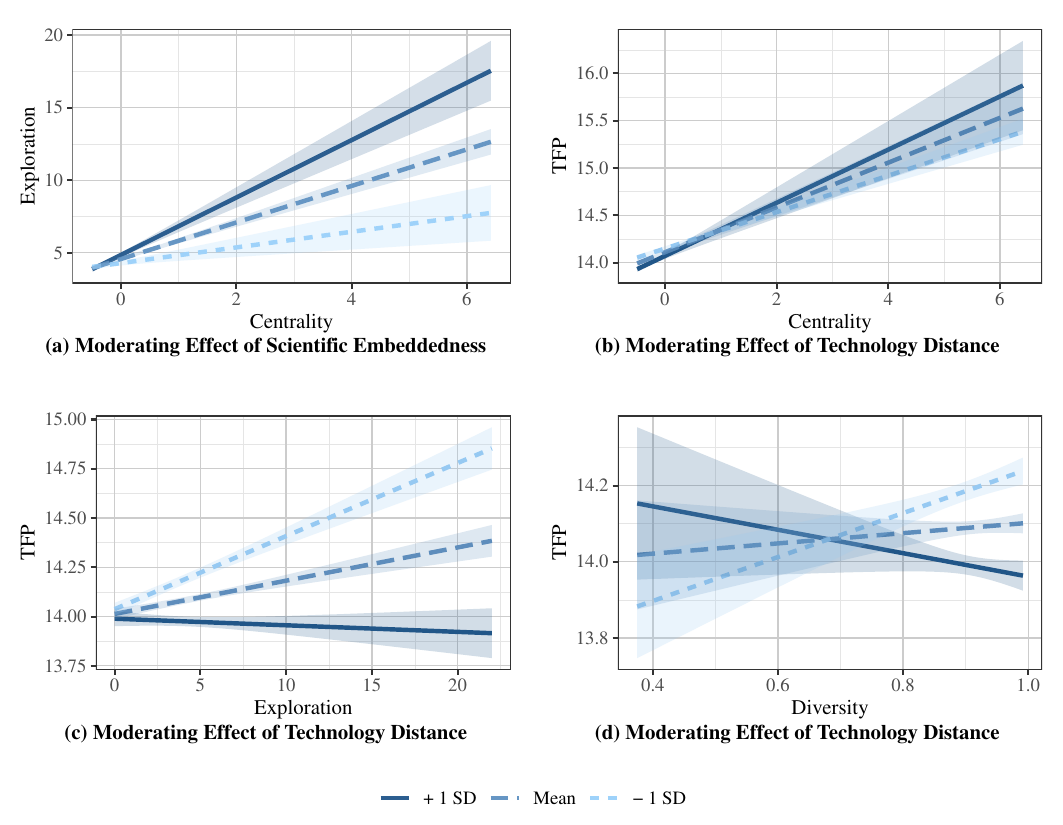}
\caption{Moderating Effects of Scientific Embeddedness and Technological Distance. 
Panel (a) illustrates the moderating role of Scientific Embeddedness in the relationship between Centrality and Exploration.
Panel (b), (c) and (d) present the moderating role of Technological Distance in the Centrality-TFP, Exploration-TFP and Diversity-TFP relationships, respectively.}
\label{Fig_moderating}
\end{figure}

Firms' innovation behavior driven by network centrality unfolds within specific cognitive and technological contexts. Whether central firms translate their advantageous network positions into productivity gains depends not only on their innovation strategies but also on the environments in which these strategies are deployed. In particular, firms differ in their degree of scientific embeddedness and in the technological distance involved in their innovation activities. These contextual conditions may shape both firms' innovation behaviors and the productivity consequences of those behaviors. We therefore examine whether scientific embeddedness and technological distance moderate the relationships linking network centrality, technological exploration, technological diversity, and firm productivity. Table~\ref{Table_Moderation} reports the estimation results, and Fig.~\ref{Fig_moderating} presents the corresponding marginal relationships.

We first examine the moderating role of scientific embeddedness. Models (1)–(3) present the results. Model (1) shows that network centrality is positively associated with firm productivity ($\beta = 0.190$, $p < 0.01$). However, the interaction between centrality and scientific embeddedness is not statistically significant ($\beta = -0.083$, $p > 0.1$), suggesting that scientific embeddedness does not directly amplify the productivity returns of central network positions. The moderating effect becomes evident when examining firms' innovation behavior. Model (2) reveals a strong positive interaction between centrality and scientific embeddedness on technological exploration ($\beta = 2.387$, $p < 0.01$). The marginal relationship displayed in Fig.~\ref{Fig_moderating}(a) shows a clear divergence between firms with high and low levels of scientific embeddedness. For firms that are deeply embedded in scientific knowledge, network centrality strongly increases the likelihood of entering new technological domains. By contrast, for firms with weak scientific linkages, the relationship between centrality and exploration is substantially attenuated. This pattern suggests that scientific engagement enhances firms' absorptive capacity, enabling central firms to better interpret heterogeneous knowledge signals circulating in innovation networks and to identify opportunities for distant technological search. Model (3) shows that the interaction between centrality and scientific embeddedness is not significant for technological diversity. This indicates that scientific embeddedness primarily shapes the direction of technological search rather than the breadth of firms' technological portfolios.

We next turn to technological distance, which reflects the cognitive complexity and recombination challenges associated with innovation across technologically distant domains. Models (4)--(6) examine how technological distance conditions the productivity consequences of network position and innovation strategies.

Model (4) shows a positive and statistically significant interaction between network centrality and technological distance ($\beta = 0.465$, $p < 0.01$). The marginal relationship plotted in Fig.~\ref{Fig_moderating}(b) indicates that the productivity advantage of central network positions becomes stronger as technological distance increases. In other words, when innovation spans more distant technological domains, occupying a central position in the patent network becomes increasingly valuable for sustaining productivity.

Models (5) and (6) reveal a striking pattern regarding firms' innovation strategies. Model (5) shows a negative interaction between technological exploration and technological distance ($\beta = -0.077$, $p < 0.01$). The marginal effects in Fig.~\ref{Fig_moderating}(c) indicate that exploration enhances productivity when technological distance is relatively low. However, as technological distance increases, the productivity returns to exploration gradually decline and eventually turn negative. A similar pattern emerges for technological diversity. Model (6) shows that the interaction between diversity and technological distance is significantly negative ($\beta = -1.961$, $p < 0.01$). While maintaining a diverse technological portfolio generally contributes to higher productivity, the magnitude of this benefit diminishes as technological distance increases. When innovation activities involve highly distant technological domains, the productivity advantage of diversification largely disappears.

Taken together, these results reveal an important tension in distant technological search. Exploration and diversification are often viewed as engines of long-term innovation, yet their short-term productivity returns are highly sensitive to technological distance. When firms attempt to recombine knowledge across distant technological domains, coordination costs and cognitive frictions increase substantially, which erodes the productivity benefits of exploratory innovation strategies. Network centrality plays a critical role in mitigating these challenges. Rather than directly improving innovation outcomes, central positions provide structural advantages that help firms cope with the complexity of distant technological search. Scientific embeddedness, by contrast, primarily operates as a behavioral catalyst that encourages central firms to initiate exploratory search, without directly strengthening the productivity consequences of those strategies.

\begin{table}[htb!]
\centering
\renewcommand\arraystretch{0.8}
\setlength{\tabcolsep}{2pt}
\caption{Moderating effects of scientific embeddedness and technological distance}
\label{Table_Moderation}
\begin{tabular*}{\linewidth}{@{\extracolsep{\fill}}lcccccc}
\toprule
 & (1) & (2) & (3) & (4) & (5) & (6) \\
\cmidrule(lr){2-7}
 & $TFP$ & $Exploration$ & $Diversity$ & $TFP$ & $TFP$ & $TFP$ \\
\midrule
$Centrality$ 
 & 0.190$^{***}$ & 0.600$^{***}$ & 0.005 & 0.051 &  &  \\
 & (0.022) & (0.181) & (0.003) & (0.033) &  &  \\
$Exploration$ 
 &  &  &  &  & 0.056$^{***}$ &  \\
 &  &  &  &  & (0.009) &  \\
$Diversity$ 
 &  &  &  &  &  & 1.692$^{***}$ \\
 &  &  &  &  &  & (0.379) \\
$SciEmbed$ 
 & $-$0.095$^{**}$ & 0.913$^{***}$ & 0.007 &  &  &  \\
 & (0.044) & (0.348) & (0.006) &  &  &  \\
$TechDist$ 
 &  &  &  & $-$0.058 & $-$0.137 & 1.324$^{**}$ \\
 &  &  &  & (0.073) & (0.086) & (0.592) \\
$Centrality$$\times$$SciEmbed$ 
 & $-$0.083 & 2.387$^{***}$ & 0.011 &  &  &  \\
 & (0.103) & (0.880) & (0.013) &  &  &  \\
$Centrality$$\times$$TechDist$ 
 &  &  &  & 0.465$^{***}$ &  &  \\
 &  &  &  & (0.113) &  &  \\
$Exploration$$\times$$TechDist$ 
 &  &  &  &  & $-$0.077$^{***}$ &  \\
 &  &  &  &  & (0.017) &  \\
$Diversity$$\times$$TechDist$ 
 &  &  &  &  &  & $-$1.961$^{***}$ \\
 &  &  &  &  &  & (0.672) \\
$Age$ 
 & 0.045 & $-$0.824$^{***}$ & 0.014$^{**}$ & 0.034 & 0.102$^{*}$ & 0.067 \\
 & (0.060) & (0.289) & (0.007) & (0.060) & (0.062) & (0.062) \\
$Labor$ 
 & 0.045$^{*}$ & 0.722$^{***}$ & 0.009$^{***}$ & 0.038 & 0.047$^{*}$ & 0.068$^{***}$ \\
 & (0.024) & (0.105) & (0.003) & (0.024) & (0.025) & (0.025) \\
$Capital$ 
 & 0.072$^{***}$ & 0.352$^{***}$ & 0.007$^{***}$ & 0.070$^{***}$ & 0.083$^{***}$ & 0.090$^{***}$ \\
 & (0.020) & (0.083) & (0.002) & (0.020) & (0.020) & (0.021) \\
$RD$ 
 & $-$3.166$^{***}$ & 1.739 & 0.118$^{***}$ & $-$3.293$^{***}$ & $-$3.116$^{***}$ & $-$3.195$^{***}$ \\
 & (0.406) & (1.584) & (0.038) & (0.403) & (0.404) & (0.398) \\
$ROA$ 
 & 0.018$^{*}$ & 0.070$^{**}$ & $-$0.000 & 0.018$^{*}$ & 0.018$^{*}$ & 0.018$^{*}$ \\
 & (0.010) & (0.028) & (0.000) & (0.009) & (0.009) & (0.010) \\
$Lev$ 
 & 0.557$^{***}$ & 0.986$^{**}$ & 0.044$^{***}$ & 0.554$^{***}$ & 0.545$^{***}$ & 0.516$^{***}$ \\
 & (0.103) & (0.453) & (0.012) & (0.102) & (0.102) & (0.105) \\
Constant 
 & 12.084$^{***}$ & $-$6.791$^{***}$ & 0.621$^{***}$ & 12.281$^{***}$ & 11.616$^{***}$ & 10.138$^{***}$ \\
 & (0.371) & (1.561) & (0.040) & (0.370) & (0.364) & (0.463) \\
\midrule
Year FE  & YES & YES & YES & YES & YES & YES \\
Ind FE   & YES & YES & YES & YES & YES & YES \\
City FE  & YES & YES & YES & YES & YES & YES \\
\midrule
Observations & 5,121 & 5,121 & 5,121 & 5,110 & 5,110 & 5,110 \\
$R^2$-adj & 0.609 & 0.400 & 0.396 & 0.614 & 0.604 & 0.597 \\
\bottomrule
\end{tabular*}
\begin{flushleft}
\footnotesize
\textit{Notes}: This table reports moderating effects of scientific embeddedness and technological distance in the relationship between network centrality, technological exploration, diversity, and firm productivity.
Standard errors clustered at the firm level are reported in parentheses.
$^{***}$, $^{**}$, and $^{*}$ denote significance at the 1\%, 5\%, and 10\% levels, respectively.
\end{flushleft}
\end{table}

\section{Discussion and conclusion}
\label{Sec_Discussion}
This study examines the behavioral mechanisms linking innovation network centrality to firms' total factor productivity. Our findings suggest that network positions do not merely serve as passive conduits for knowledge accumulation. Instead, firms occupying central positions systematically reorient their innovation activities along two dimensions. First, central firms increase exploratory entry into new technological domains. Second, they broaden the structure of their technological portfolios by engaging with a more diverse set of knowledge areas. These results indicate that productivity advantages associated with network centrality emerge not from structural position alone but from the strategic reconfiguration of technological search enabled by that position.

Building on this behavioral shift, we show how network advantages are translated into productivity outcomes through a set of mediated and moderated pathways. Central network positions promote both exploratory entry into new technological domains and greater technological diversity in firms' innovation portfolios. Together these behavioral adjustments represent a dual reorientation of technological search that reshapes how firms allocate innovation efforts across unfamiliar and heterogeneous knowledge domains. The effectiveness of this reorientation depends on critical contextual conditions. Scientific embeddedness strengthens the influence of network centrality on exploratory entry and highlights the role of scientific knowledge in guiding opportunity recognition during the early stages of technological search. In contrast, technological distance conditions the economic returns to innovation strategies. As technological distance increases, the productivity benefits of both exploration and technological diversity diminish, reflecting the growing organizational challenges involved in integrating distant knowledge domains. By distinguishing these contextual mechanisms, our study demonstrates that the value of innovation network centrality depends on firms' ability to strategically leverage their network positions to reshape both the direction and the structure of technological search.

We acknowledge several limitations that open avenues for future research. Patent data provide a valuable lens on technological activities but may underrepresent innovation in service sectors or industries with low patenting propensity. Future studies could incorporate alternative indicators such as research and development project portfolios or new product introductions. Our focus on scientific embeddedness and technological distance as contextual moderators leaves open the potential influence of other factors, including organizational learning routines, managerial cognition, or industry technological turbulence. Although our empirical design establishes associations consistent with the proposed mechanisms, future research could adopt quasi-experimental or longitudinal designs to strengthen causal inference.

For firm managers, our results suggest that the value of occupying a central network position lies in its capacity to support the strategic rebalancing of innovation portfolios. Central firms appear better positioned not only to explore emerging technological domains but also to maintain broader technological portfolios that span heterogeneous knowledge areas. These capabilities allow firms to redirect innovation efforts away from increasingly saturated technological trajectories and toward new opportunities while maintaining sufficient diversity to sustain long-term adaptability. In this sense, network centrality provides firms with the informational and relational resources necessary to restructure technological search in ways that enhance productivity growth.

For policymakers, the findings highlight the importance of fostering institutional environments that enable firms to translate network advantages into high-quality innovation search. Rather than focusing exclusively on expanding collaborative ties or increasing patent counts, policy frameworks should emphasize mechanisms that support experimentation across technological boundaries and reduce the uncertainty associated with exploratory innovation. Improving the accessibility and interpretability of scientific knowledge can function as a form of public infrastructure that lowers cognitive barriers to cross-domain search. In addition, innovation policies that mitigate the risks associated with technologically distant recombination can help firms capture the productivity benefits of diversified and exploratory innovation strategies. By enabling firms to navigate technological opportunities while managing integration challenges, such policies can promote more sustainable and efficiency-oriented technological progress.

\section*{Acknowledgment}
This research was supported by the National Social Science Foundation of China (23XJL006), the Hainan Provincial Natural Science Foundation High-Level Talent Project (724RC496), and the Hainan Provincial Natural Science Foundation Youth Fund Project (722QN296).

\section*{Disclosure statement}

No potential conflict of interest was reported by the authors.

\section*{Data availability}

Data will be made available on request.

\clearpage

\bibliographystyle{elsarticle-harv}
\bibliography{Reference}

\end{document}